\documentclass[12pt]{article}

\usepackage{latexsym}

\def\cH{{\cal H}}

\def\tr{{\rm tr}}
\def\ket#1{\mid~\!\!\!{#1}~\!\!\rangle}
\def\bra#1{\langle~\!\!{#1}~\!\!\!\mid}
\def\IF{if and only if }

\def\QMl{quantum-mechanical }

\def\CC{calibration condition }
\def\cc{calibration condition}

\def\ON{orthonormal }

\def\${\enskip$}
\def\M{measurement }
\def\m{measurement}

\begin{document}

{\bf\noindent \large On quantum subsystem  measurement}
\vspace{0.5cm}

{\bf \noindent F. Herbut}\\
\vspace{0.3cm}
\rm

\noindent {\bf  Abstract.} It is assumed that an arbitrary composite bipartite pure state in which the two subsystems are entangled is given, and it is investigated how the entanglement transmits the influence of \M on only one of the subsystems to the state of the opposite subsystem. It is shown that any exact subsystem \M  has the same influence as ideal \M on the opposite subsystem. In particular, the distant effect of subsystem \M of a twin observable, i. e., so-called 'distant \m ', is always ideal \M on the distant subsystem no matter how intricate the direct exact \M on the opposite subsystem is.\\
\vspace{0.3cm}

\noindent {\bf Keywords} Entanglement in \m . Measurement effects due to entanglement. Unitary \m . Basic dynamics.\\

\vspace{0.5cm}
\normalsize \rm

{\bf \noindent 1 Introduction}\\

\noindent The present article investigates some implications of defining the measuring process by a {\bf unitary operator} that incorporates the interaction between object and measuring instrument. One deals with so-called {\bf nonselective \m }, i. e., \M short of collapse (if done on an ensemble, this contains all the results). So-called {\bf selective \M } is \M with collapse, when one result is considered (the subensemble of this result is selected). The mechanism of collapse is known to lie outside unitary dynamics \cite{vNeum}. It will not be considered in this study. Most interpretations of collapse are in agreement with the \QMl formalism, which implies the unitary \M dynamics presented.

In the literature by \M one usually means selective \m . In this article we mean by \M nonselective \M  unless otherwise stated.

{\footnotesize \rm \noindent
\rule[0mm]{4.62cm}{0.5mm}

\noindent F. Herbut (mail)\\
Serbian Academy of Sciences and Arts,
Knez Mihajlova 35, 11000 Belgrade,
Serbia\\
e-mail: fedorh@sanu.ac.rs}
\pagebreak

The terms 'pure state' and 'state vector' (vector of norm one) will be used interchangeably; and so will 'state' and 'density operator', 'observable' and 'Hermitian operator' (with a purely discrete spectrum), event' and 'projector' throughout the paper.  The subsystem over which a partial trace is taken will be denoted by index (or indices). Total traces go without indices.\\

Let subsystem A be the object of \m , and let $$O_A=\sum_ko_kE_A^k,\quad k\not= k'\enskip
\Rightarrow\enskip o_k\not= o_{k'}\eqno{(1)}$$ be {\bf the measured observable} (Hermitian operator with a purely discrete, finite or infinite, spectrum) {\bf in its unique spectral form}. By 'uniqueness' is meant the non-repetition of the eigenvalues \$\{o_k:\forall k\}\$ in (1)). Henceforth, we always mean by 'spectral form' the unique one unless otherwise stated.

Naturally, also the completeness relation \$\sum_kE_A^k=I_A\$, \$I_A\$ being the identity operator for subsystem A, is valid. Let, further, subsystem B be the measuring instrument equipped with a pointer observable
$$P_B=\sum_kp_kF_B^k,\eqno{(2)}$$ also in its spectral form. The completeness relation \$\sum_kF_B^k=I_B\$ is valid too.

The measuring apparatus 'takes cognizance' of the results, eigenvalues \$o_k\$ or, equivalently, of the corresponding eigen-events \$E_A^k\$, in terms of its 'pointer positions', which are either the eigenvalues \$p_k\$ of the pointer observable or, equivalently, the eigen-events \$F_B^k\$. (This is stated more precisely below when \M is defined.)

Finally, let \$U_{AB}\$ be the unitary operator incorporating the {\bf \M interaction} and mapping any initial composite-system state vector \$\ket{\phi}_A\otimes\ket{\phi}_B^i\$ into the {\bf final state} (at the end of \M interaction): $$\ket{\Phi}_{AB}^f\equiv U_{AB}\Big(\ket{\phi}_A\ket{\phi}_B^i\Big).\eqno{(3)}$$

By \$\ket{\phi}_A\$ is denoted an arbitrary state vector of the measured system \$A\$, and \$\ket{\phi}_B^i\$ is {\bf the initial or ready-to-measure state vector} of the instrument.

We use the convention that kets and bras denote state vectors.\\

In this investigation the basic aim is to focus attention on {\bf bipartite composite systems} in some pure state \$\ket{\Phi}_{A_1A_2}\$ where \$A\equiv A_1+A_2\$ is the object of \m . We are particularly interested in {\bf subsystem \m s} on subsystem \$A_2\$, which we call the {\bf nearby} subsystem, and on its influence on the opposite, dynamically unaffected subsystem \$A_1\$, called {\bf distant} or {\bf remote}. (The terms are dynamical, not spatial.) The influence is transmitted by the {\bf entanglement} in the composite state.\\

{\bf \noindent  2 Definition and Basic Dynamical Property of Measurement}\\

\noindent {\bf Exact measurement} is defined by requiring the validity of the so-called {\bf \CC} \cite{BLM}. It reads: If the initial state of the object has a definite value of the measured observable, then the final composite-system state has the {\bf corresponding} definite value of the pointer observable . 'Corresponding' we write as 'having the same index value' (cf (1) and (2)).\\

Since approximate \m s are not studied in this article, henceforth we drop the term 'exact'.\\

All \QMl relations have a statistical meaning and are tested on ensembles of equally prepared systems. The precise {\bf statistical form} of the {\bf \CC} is expressed in terms of the usual {\bf probability formulae}:
$$\forall k:\enskip
\bra{\phi}_AE_A^k\ket{\phi}_A=1\quad\Rightarrow\quad \bra{\Phi}_{AB}^fF_B^k\ket{\Phi}_{AB}^f=1,\eqno{(4)}$$ where \$\Rightarrow\$ denotes logical implication, and the final state \$\ket{\Phi}_{AB}^f\$ is given by (3).

To derive an equivalent, more practical, form of (4), we need a useful general and known, but perhaps not well known, auxiliary {\bf claim} (proved in Appendix A for the reader's convenience).\\

An event \$E\$ is certain, i. e., has probability one, in a pure state \$\ket{\psi}\$ \IF the former, acting on the latter, does not change it:
$$\bra{\psi}E\ket{\psi}=1 \enskip\Leftrightarrow\enskip E\ket{\psi}=\ket{\psi}.\eqno{(5)}$$ (The symbol "$\Leftrightarrow$" denotes logical implication in both directions.)\\

Equivalence (5) makes it obvious that the calibration condition can be equivalently expressed in the more practical form:
$$\forall k:\qquad\ket{\phi}_A=E_A^k\ket{\phi}_A
\quad\Rightarrow\quad\ket{\Phi}_{AB}^f=
F_B^k\ket{\Phi}_{AB}^f\eqno{(6)}$$ (cf (1)-(3)).\\

Now we state and prove the {\bf basic  dynamical property} of \m . Actually, it is a necessary and sufficient condition for the \cc , or otherwise put, it is another definition of \m .(We call it "dynamical" because it involves the unitary evolution operator \$U_{AB}\$ explicitly.) The {\bf claim} goes as follows.\\

One has  \M  {\bf \IF } $$\forall\ket{\phi}_A,\enskip\forall k:\enskip\Big(F_B^kU_{AB}\Big)\Big(\ket{\phi}_A
\ket{\phi}_B^i\Big)= \Big(U_{AB}E_A^k\Big)\Big(\ket{\phi}_A\ket{\phi}_B^i\Big) \eqno{(7)}$$ is valid.\\

One {\it proves necessity} as follows. The completeness relation \$\sum_{k'}E_A^{k'}=I_A\$, repeated use of the \CC (6), and orthogonality and idempotency of the \$F_B^k\$ projectors enable one to write for each \$k\$ value (we shall put \$\times\$ after a number whenever a term in an expansion begins by that number): $$F_B^kU_{AB}\ket{\phi}_A\ket{\phi}_B^i=$$ $$\sum_{k'}||E_A^{k'}\ket{\phi}_A||\times F_B^k
U_{AB}\Big( E_A^{k'}\ket{\phi}_A\Big/ ||E_A^{k'}\ket{\phi}_A||\Big)
\ket{\phi}_B^i=$$ $$\sum_{k'}||E_A^{k'}\ket{\phi}_A||\times F_B^k
\mathbf{F_B^{k'}}U_{AB}\Big( E_A^{k'}\ket{\phi}_A\Big/ ||E_A^{k'}\ket{\phi}_A||\Big)
\ket{\phi}_B^i=$$ $$||E_A^k\ket{\phi}_A||\times F_B^kU_{AB}\Big( E_A^k\ket{\phi}_A\Big/ ||E_A^k\ket{\phi}_A||\Big)
\ket{\phi}_B^i.$$ Finally, on account of (6) the auxiliary claim (5) allows one to omit \$F_B^k\$, so that, after cancelation , one obtains: $$lhs=
U_{AB}E_A^k\ket{\phi}_A\ket{\phi}_B^i.$$

To {\it prove sufficiency}, let $$\Big( U_{AB}E_A^k\Big) \Big(\ket{\phi}_A\ket{\phi}_B^i\Big)= \Big(F_B^kU_{AB}\Big)
\Big(\ket{\phi}_A\ket{\phi}_B^i\Big)$$ be valid for all \$k\$ values, and let \$\ket{\phi}_A=E_A^{k'}\ket{\phi}_A\$ be satisfied for a fixed value  \$k\equiv k'\$. Then, one has in particular $$\Big( U_{AB}E_A^{k'}\Big) \Big(\ket{\phi}_A\ket{\phi}_B^i\Big)= \Big(F_B^{k'}U_{AB}\Big)
\Big(\ket{\phi}_A\ket{\phi}_B^i\Big).$$ One can here omit \$E_A^{k'}\$ due to the assumed definite value using (5), and thus the \CC (6) is obtained. {\it This ends the proof.}\\

{\bf \noindent 3 Subsystem Measurement in Composite State}\\

In this section we assume that an arbitrary composite bipartite system \$A\equiv A_1+A_2\$ in an arbitrary pure state \$\ket{\phi}_{A_1,A_2}\$  and an arbitrary  subsystem observable \$O_{A_2}=\sum_ko_kE_{A_2}^k\$ for the nearby subsystem are given. We investigate the consequences of the basic dynamical characterization of \M (7) for this case to find out how entanglement transmits the subsystem \M dynamics on the nearby subsystem \$A_2\$ onto the state of the remote opposite subsystem \$A_1\$.

To begin with, it is known that any unitary change to subsystem \$A_2\$, with or without an ancilla \$A_3\$, does not have any influence on the state of subsystem \$A_1\$.

More precisely, the {\bf claim} is that, if there is no interaction between subsystems \$A_1\$ and \$A_2+A_3\$, i. e., if the composite unitary evolution operator can be factorized \$U_{A_1,A_2,A_3}=U_{A_1}\otimes U_{A_2,A_3}\$, then the {\bf final remote subsystem state} reads $$\rho_{A_1}^f\equiv \tr_{A_2,A_3}\Big(U_{A_1,A_2,A_3}\ket{\phi}_{A_1,A_2,A_3}  \bra{\phi}_{A_1,A_2,A_3}U_{A_1,A_2,A_3}^{\dag}\Big)= U_{A_1}\rho_{A_1}^iU_{A_1}^{\dag},\eqno{(8)}$$ where \$\rho_{A_1}^i\equiv \tr_{A_2,A_3}(\ket{\phi}_{A_1,A_2,A_3}  \bra{\phi}_{A_1,A_2,A_3})\$ is the initial state of subsystem \$A_1\$ in the composite-system state \$\ket{\phi}_{A_1,A_2,A_3}= \ket{\phi}_{A_1,A_2}\ket{\phi}_{A_3}\$.

Note that what makes the ancilla \$A_3\$ an auxiliary system is the fact that it is initially uncorrelated with the system \$A_1+A_2\$ that is considered. Further, one should note that if there is no interaction with the ancilla, then the ancilla evolves independently, and it can be disregarded.

Though claim (8) is known, for the reason of completeness, we sketch the proof. But for this we need a general auxiliary claim, which will be referred to as the {\bf 'under-the-partial-trace commutativity'} (it will be used again below). It reads:

$$O_A\equiv\tr_B\Big(\mathbf{Y_B}X_{AB}\Big)=
\tr_B\Big(X_{AB}\mathbf{Y_B}\Big) ,\eqno{(9)}$$ where \$Y_B\$ and \$X_{AB}\$ are arbitrary subsystem and composite-system operators respectively. This general {\bf claim} is proved in Appendix B.\\

{\it Proof} for claim (8) follows immediately from the definition in (8) when one takes into account the facts (i) that opposite-subsystem operators can be taken out of the partial trace preserving the order of the operators (\$U_{A_1}\$ and \$U_{A_1}^{\dag}\$ in this case), (ii) that one has the under-the-partial-trace commutativity (9), which concerns \$U_{A_2,A_3}\$ with the rest, and finally, (iii) that a unitary operator (\$U_{A_2,A_3}\$ in this case) multiplied by its inverse gives the identity operator. {\it This ends the proof.}\\

Since a \M instrument \$B\$ qualifies for an ancilla (cf (3)), though its role is far from auxiliary, it is clear from claim (8) that {\bf nonselective \m } of any nearby subsystem observable \$O_{A_2}\$ in any pure state of a composite system \$A_1+A_2\$ {\bf cannot influence the state} of the distant subsystem \$A_1\$.\\

Next, we are interested in {\bf selective subsystem \m }. The {\bf general claim}, a consequence of the basic dynamical relation (7), goes as follows.

{\bf Selective \M does, in general, influence} the state of the remote subsystem \$A_1\$. More precisely, if
a nearby-subsystem observable \$O_{A_2}=\sum_ko_kE_{A_2}^k\$ is measured selectively with the result \$o_k\$ in a bipartite pure state \$\ket{\phi}_{A_1,A_2}\$ in which one has positive probability \$\bra{\phi}_{A_1,A_2}E_{A_2}^k\ket{\phi}_{A_1,A_2}>0\$, then the {\bf final selective distant-subsystem state} $$\rho_{A_1}^{f,k}\equiv$$ $$\tr_{A_2,B}\Big[\Big(
F_B^k\ket{\Phi}_{A_1,A_2,B}^f\Big/
||F_B^k\ket{\Phi}_{A_1,A_2,B}^f||\Big)\Big(
\bra{\Phi}_{A_1,A_2,B}^fF_B^k\Big/
||F_B^k\ket{\Phi}_{A_1,A_2,B}^f||\Big)\Big]\eqno{(10)}$$  has the form: $$\rho_{A_1}^{f,k}= U_{A_1}\Big(\rho_{A_1}(E_{A_2}^k)\Big)U_{A_1}^{\dag},
\eqno{(11a)}$$ where by $$\rho_{A_1}(G_{A_2})\equiv\tr_{A_2}\Big(
(\ket{\phi}_{A_1,A_2}\bra{\phi}_{A_1,A_2})G_{A_2}
\Big)\Big/ \tr\Big((\ket{\phi}_{A_1,A_2}
\bra{\phi}_{A_1,A_2})G_{A_2}\Big)\eqno{(11b)}$$ (\$G_{A_2}\$ being any projector in the state space \$\cH_{A_2}\$) is denoted the {\bf conditional state} of the remote subsystem \$A_1\$ under the condition of the occurrence of the event \$G_{A_2}\$ in the composite-system state \$\ket{\phi}_{A_1,A_2}\$, and  \$U_{A_1}\$ is the unitary evolution operator of the remote subsystem.\\

To {\it prove} (11a), we evaluate \$\rho_{A_1}^{f,k}\$ from its definition (10). By this we utilize the following equalities, which are a consequence of  (7) and (3), of the fact that a unitary operator does not change the norm, and finally of the fact that the norm of a tensor product is the product of the norms. $$||F_B^k\ket{\Phi}_{A_1,A_2,B}^f||=
||E_{A_2}\ket{\phi}_{A_1,A_2}||=$$ $$
\Big(
\bra{\phi}_{A_1,A_2}E_{A_2}^k\ket{\phi}_{A_1,A_2}\Big)^{1/2}=
\Big[\tr\Big((\ket{\phi}_{A_1,A_2}\bra{\phi}_{A_1,A_2})
E_{A_2}^k\Big)\Big]^{1/2}.\eqno{(12)}$$ Besides (12), we take again resort to (7), take into account the partial-trace property that opposite-subsystem operators can be taken out of the partial trace (preserving the order of the  operators as factors), as well as the 'under-the-partial-trace commutativity' (9) twice:

 $$\rho_{A_1}^{f,k}=\Big(
\bra{\phi}_{A_1,A_2}E_{A_2}^k\ket{\phi}_{A_1,A_2}\Big)^{-1}\times
$$ $$\tr_{A_2,B} \Big[\Big(U_{A_1}U_{A_2,B}E_{A_2}^k(\ket{\phi}_{A_1,A_2}
\ket{\phi}_B^i)\Big) \Big((\bra{\phi}_{A_1,A_2}
\bra{\phi}_B^i)E_{A_2}^kU_{A_1}^{\dag}U_{A_2,B}^{\dag}\Big)
\Big]=$$

$$U_{A_1}\Big\{\tr_{A_2B}\Big[\Big(E_{A_2}^k(\ket{\phi}_{A_1,A_2}
\ket{\phi}_B^i
\bra{\phi}_{A_1,A_2}\bra{\phi}_B^i)E_{A_2}^k\Big) \Big(U_{A_2B}^{\dag}
U_{A_2B}\Big)\Big]\Big\}U_{A_1}^{\dag}\Big/$$
$$\tr\Big((\ket{\phi}_{A_1,A_2}
\bra{\phi}_{A_1,A_2})E_{A_2}^k\Big)=$$

$$U_{A_1}\Big\{\Big[\tr_{A_2}\Big(E_{A_2}^k(\ket{\phi}_{A_1,A_2}
\bra{\phi}_{A_1,A_2}E_{A_2}^k\Big)\Big] \Big[\tr_B
\Big(\ket{\phi}_B^i\bra{\phi}_B^i\Big)\Big]\Big\}U_{A_1}^{\dag}\Big/$$
$$\tr\Big((\ket{\phi}_{A_1,A_2}
\bra{\phi}_{A_1,A_2})E_{A_2}^k\Big)=$$

$$U_{A_1}\Big[\tr_{A_2}\Big((\ket{\phi}_{A_1,A_2}
\bra{\phi}_{A_1,A_2})E_{A_2}^k\Big)\Big]U_{A_1}^{\dag}\Big/
\tr\Big((\ket{\phi}_{A_1,A_2}
\bra{\phi}_{A_1,A_2})E_{A_2}^k\Big)=$$ $$U_{A_1}
\rho_{A_1}(E_{A_2}^k)U_{A_1}^{\dag}.$$ {\it This ends the proof.}\\

It is important to note that claim (11a) implies that it is irrelevant what kind of \M is performed on the nearby subsystem, the effect on the distant subsystem is {\bf one and the same}, and the influence of the \M goes only in terms of  the {\bf eigen-projectors} of the measured observable. Another way to express this fact is to say that any \M on the nearby subsystem acts on the distant subsystem equally as the simplest, i. e., {\bf ideal \m }.\\

Consistency of no change in nonselective \M on the one hand, and of the evaluated change in selective \M on the other, i. e., of (8) and (11a), is seen in the following decomposition. $$\rho_{A_1}^i=\sum_k\Big(
\bra{\phi}_{A_1,A_2}E_{A_2}^k\ket{\phi}_{A_1,A_2}\Big)
\rho_{A_1}(E_{A_2}^k).\eqno{(13)}$$

To {\it prove} decomposition (13), we make use of the completeness relation \$\sum_kE_{A_2}^k=I_{A_2}\$ and of (12): $$\rho_{A_1}^i=\sum_k\Big(
\bra{\phi}_{A_1,A_2}E_{A_2}^k\ket{\phi}_{A_1,A_2}\Big)\times
\Big\{\tr_{A_2}\Big(\ket{\phi}_{A_1,A_2}\bra{\phi}_{A_1,A_2}
E_{A_2}^k\Big)\Big/$$ $$\Big[\tr\Big(\ket{\phi}_{A_1,A_2}
\bra{\phi}_{A_1,A_2}E_{A_2}^k\Big)\Big]\Big\}.$$ In view of (11b), {\it this ends the proof.}\\

One should note that any orthogonal projector decomposition of the identity operator \$I_{A_2}\$ induces likewise a decomposition of \$\rho_{A_1}^i\$ (displays the density operator as an improper mixture \cite{D'Esp}). For the \M of \$O_{A_2}=\sum_ko_kE_{A_2}^k\$ one of this mixtures, particularly (13), is relevant.

Relation (11a) tells us that all that selective nearby-subsystem \M with the result \$o_k\$  accomplishes on the remote subsystem is that it picks the state \$\rho_{A_1}(E_{A_2}^k)\$ in the corresponding mixture (13). In view of (8), the state \$\rho_{A_1}(E_{A_2}^k)\$ then evolves according to the dynamics of the remote subsystem with no regard to the chosen \M on the nearby system.

This insight might be useful for any theory of collapse, i. e., of selective \m .

\pagebreak

{\bf \noindent 4 Subsystem Measurement of Twin Observable; Distant Measurement}\\

\noindent Now we assume that, for a given bipartite pure state \$\ket{\phi}_{A_1,A_2}\$, a pair of (opposite subsystem) {\bf twin observables} \$O_{A_1}\$ and \$O_{A_2}\$ are given. By definition, they can be written as
$$O_{A_q}=\sum_ko_k^{(q)}E_{A_q}^k+O_{A_q}',\quad
q=1,2\enskip,\eqno{(14a,b)}$$ where the the sums are written as unique spectral forms, and also $$\forall k:\quad
E_{A_1}^k\ket{\phi}_{A_1,A_2}=E_{A_2}^k
\ket{\phi}_{A_1,A_2};\eqno{(14c)}$$ $$ O_{A_q}'\ket{\phi}_{A_1,A_2}=0,\enskip q=1,2\eqno{(14d)}$$ are valid (cf \cite{FHPR}).\\

The following {\bf claim}  holds true. If only \$O_{A_2}\$ of the above pair of {\bf twin observables} is measured selectively on the nearby subsystem with the result \$o_k^{(2)}\$, then the final state of the remote subsystem is $$ \rho_{A_1}^{f,k}=U_{A_1}\Big\{E_{A_1}^k\rho_{A_1}^i
E_{A_1}^k\Big/\Big[\tr\Big(\rho_{A_1}^i
E_{A_1}^k\Big)\Big]\Big\}
U_{A_1}^{\dag},\eqno{(15)}$$ and this is valid for every value of \$k\$.\\

To {\it prove} claim (15), we make use of (11b), of idempotency, of under-the-partial-trace commutativity, of the twin-observables definition (14c), and finally of the possibility to take out opposite-subsystem operators from the partial trace: $$\rho_{A_1}(E_{A_2}^k)\equiv\tr_{A_2}\Big(
(\ket{\phi}_{A_1,A_2}
\bra{\phi}_{A_1,A_2})E_{A_2}^k\Big)\Big/\Big[\tr\Big(
(\ket{\phi}_{A_1,A_2}
\bra{\phi}_{A_1,A_2})E_{A_2}^k\Big)\Big]=$$ $$ \tr_{A_2}\Big((E_{A_2}^k\ket{\phi}_{A_1,A_2})
(\bra{\phi}_{A_1,A_2}E_{A_2}^k)\Big)\Big/
\Big[\tr\Big((
\ket{\phi}_{A_1,A_2}
\bra{\phi}_{A_1,A_2})E_{A_2}^k\Big)\Big]=$$ $$ \tr_{A_2}\Big((\mathbf{E_{A_1}^k}\ket{\phi}_{A_1,A_2})
(\bra{\phi}_{A_1,A_2}\mathbf{E_{A_1}^k})\Big)\Big/
\Big\{\tr_{A_1}\Big[\tr_{A_2}\Big(
\ket{\phi}_{A_1,A_2}
\bra{\phi}_{A_1,A_2}\Big)\Big]\mathbf{E_{A_1}^k}\Big\}=$$
$$E_{A_1}^k\rho_{A_1}^iE_{A_1}^k\Big/\Big[\tr\Big(
\rho_{A_1}^iE_{A_1}^k\Big)\Big].$$ In view of (11a), {\it this ends the proof.}\\

The change of state $$\rho_{A_1}^i\quad\rightarrow\quad
E_{A_1}^k\rho_{A_1}^iE_{A_1}^k\Big/\tr(\rho_{A_1}^iE_{A_1}^k)
\eqno{(16a)}$$ is the well-known L\"uders selective change-of-state formula (cf \cite{Lud}, \cite{LudMessiah}, \cite{LudLaloe}), which characterizes {\bf ideal selective \m  }.

One should note that \$\tr(\rho_{A_1}^iE_{A_1}^k)=\bra{\phi}_{A_1,A_2}
\mathbf{E_{A_2}^k}
\ket{\phi}_{A_1,A_2}\$ (cf (12)) is the probability of the result \$o_k^{(2)}\$. Hence, the nonselective version of the same subsystem \M on the nearby subsystem \$A_2\$ gives rise to $$\sum_k\tr(\rho_{A_1}^iE_{A_1}^k)\Big[E_{A_1}^k
\rho_{A_1}^iE_{A_1}^k\Big/\tr(\rho_{A_1}^iE_{A_1}^k)\Big]=
\sum_kE_{A_1}^k\rho_{A_1}^iE_{A_1}^k.\eqno{(16b)}$$

This is not distinct from \$\rho_{A_1}^i\$ because the completeness relation \$\sum_kE_{A_1}=I_{A_1}\$ implies
\$\rho_{A_1}^i=\sum_{k,k'}E_{A_1}^k\rho_{A_1}^iE_{A_1}^{k'}\$,
and, for \$k\not= k'\$, one has on account of the twin relation (14c), under-the-partial-trace commutativity, and orthogonality of the eigen-projectors: $$E_{A_1}^k\rho_{A_1}^iE_{A_1}^{k'}\equiv\tr_{A_2}\Big(
E_{A_1}^k\ket{\phi}_{A_1,A_2}\bra{\phi}_{A_1,A_2}E_{A_1}^{k'}
\Big)=$$ $$\tr_{A_2}\Big(
\mathbf{E_{A_2}^k}\ket{\phi}_{A_1,A_2}\bra{\phi}_{A_1,A_2}\
\mathbf{E_{A_2}^{k'}}\Big)=\tr_{A_2}\Big(
\ket{\phi}_{A_1,A_2}\bra{\phi}_{A_1,A_2}(E_{A_2}^{k'}E_{A_2}^k)
\Big)=0.$$

Naturally, the fact that nonselective subsystem \M of a twin observable on the nearby subsystem causes no change in the state of the distant subsystem is a special case of the general statement that every nearby subsystem \M behaves in this way (that is proved in claim (8)).

Result (15) can be read in the following manner: An instantaneous ideal \M of \$\mathbf{O_{A_1}}\$ appears to be performed on the initial distant-subsystem state \$\rho_{A_1}^i\$ , and then the state evolves in its unitary way till the end of the \M of \$O_{A_2}\$ on the nearby subsystem. The defining relations (11c) immediately implied this statement for ideal \M on subsystem \$A_2\$. Now, on account of the claim (11a), which covers {\bf all} \m s on the nearby subsystem, we have the general validity of the statement.

The notion of distant \m , introduced in \cite{FHMV76}, covered only the case when ideal subsystem \M was performed on the nearby subsystem and it gave rise to ideal \M on the remote subsystem (without interaction, only due to the entanglement). Since one rarely succeeds to perform ideal \M in direct interaction, the distant-\M concept was thus on feet of clay. Now the notion of {\bf distant \M } is on firm ground: Any \M of a twin observable \$\mathbf{O_{A_2}}\$ (cf (14a-d)) on the nearby subsystem brings about {\bf distant}, i. e., interaction free, {\bf ideal \M } of its twin observable \$\mathbf{O_{A_1}}\$ on the opposite, remote subsystem.\\

{\bf Appendix A. Relation of certainty in a pure state}\\

We {\it prove} now the general claim that the following equivalence is valid for a pure state \$\ket{\psi}\$ and an event \$E\$:
$$\bra{\psi}E\ket{\psi}=1\quad\Leftrightarrow\quad
\ket{\psi}=E\ket{\psi}.$$ One can write
$$\bra{\psi}E\ket{\psi}=1
\quad\Rightarrow\bra{\psi}E^c\ket{\psi}=0,$$
where \$E^c\equiv I-E\$ is the ortho-complementary projector and \$I\$ is the identity operator. Further, one has \$||E^c\ket{\psi}||=0\$,  \$E^c\ket{\psi}=0\$, and \$E\ket{\psi}=\ket{\psi}\$ as claimed.\\

{\bf Appendix B. Under-the-partial-trace commutativity}\\

We prove now the general relation
$$\tr_B\Big(Y_BX_{AB}\Big)=\tr_B\Big(X_{AB}Y_B\Big) $$ (cf (9)) by straightforward evaluation of both sides in an arbitrary pair of complete \ON bases \$\{\ket{k}_A:\forall k\}\$, \$\{\ket{n}_B:\forall n\}\$.

$$\bra{k}_Alhs\ket{k'}_A=\sum_n\bra{k}_A\bra{n}_B(Y_BX_{AB})
\ket{k'}_A\ket{n}_B=$$ $$\sum_n\sum_{k''}\sum_{n'}\bra{k}_A\bra{n}_B
(I_A\otimes Y_B)\ket{k''}_A
\ket{n'}_B\times \bra{k''}_A\bra{n'}_B(X_{AB})\ket{k'}_A\ket{n}_B=$$
$$\sum_n\sum_{n'}\bra{n}_BY_B\ket{n'}_B\times \bra{k}_A\bra{n'}_B(X_{AB})\ket{k'}_A\ket{n}_B.$$\\

$$\bra{k}_Arhs\ket{k'}_A=\sum_n\bra{k}_A\bra{n}_B(X_{AB}Y_B)
\ket{k'}_A\ket{n}_B=$$  $$\sum_n\sum_{k''}\sum_{n'}\bra{k}_A\bra{n}_BX_{AB}
\ket{k''}_A\ket{n'}_B\times \bra{k''}_A\bra{n'}_B(I_A\otimes Y_B)\ket{k'}_A\ket{n}_B=$$ $$\sum_n\sum_{n'}\bra{k}_A\bra{n}_BX_{AB}
\ket{k'}_A\ket{n'}_B\times\bra{n'}_B Y_B\ket{n}_B.$$
Finally, we exchange the order of the two factors and the two mute indices \$n\$ and \$n'\$ to obtain $$\bra{k}_Arhs\ket{k'}_A= \sum_{n'}\sum_n\bra{n}_BY_B\ket{n'}_B\times \bra{k}_A\bra{n'}_B(X_{AB})\ket{k'}_A\ket{n}_B.$$ Thus, we see that \$lhs=rhs\$ as claimed.\\

\end{document}